\documentclass[aps,prc,reprint,groupedaddress,showkeys,showpacs]{revtex4-1}
\usepackage{graphicx}% Include figure files
\usepackage{dcolumn}% Align table columns on decimal point
\usepackage{bm}%
\usepackage{amsmath}
\usepackage{capt-of}

\begin{document}

\title{Gravitational waves from phase transition of NS to QS}

\date{\today}

\author{R Prasad}
\affiliation{Indian Institute of Science Education and Research Bhopal, Bhopal, India}
\author{Ritam Mallick} 
\email{mallick@iiserb.ac.in}
\affiliation{Indian Institute of Science Education and Research Bhopal, Bhopal, India}

\begin{abstract}
In this article, we perform a 2-d simulation of combustion of neutron star (NS) to hybrid star (HS). We assume that a sudden density fluctuation at the center of the NS initiates a shock discontinuity near the center of the star. This shock discontinuity deconfines NM to 2-f QM, initiating combustion of the star. This combustion front propagates from the center to the surface converting NM to 2-f QM. This combustion stops at a radius of $6 km$ inside the star, as at this density the NM is much stable than QM. Beyond $6 km$ although the combustion stops but the shock wave propagates to the surface. We study the gravitational wave signal for such a PT of NS to HS. We find that such PT has unique GW strain of amplitude $10^{-22}$. These signals last for few tens of $\mu s$ and shows small oscillating behaviour. The power spectrum consists of peaks and at fairly high frequency range. The conversion to NS to HS has a unique signature which would help in defining the PT and the fate of the NS. 

\end{abstract}

%\pacs{47.40.Nm, 52.35.Tc, 26.60.Kp, 97.10.Cv}


\maketitle
\section{Introduction}

The two recent observation of gravitational wave (GW) from black holes merger (BHM) GW150914 \cite{abbott} 
and from a binary neutron star merger (BNSM) GW170817 \cite{abbott1} has opened a new window towards gravitational wave
astronomy and multimessenger astronomy. In the BHM GW is the only signal which otherwise would have left minimal signatures, whereas for BNSM both GW and electromagnetic radiation were detected which opens a whole new window to probe those objects which till
now has limited access. The multimessenger detection of BNSM has put a severe constraint on the nature of matter that can occur inside a neutron star (NS) \cite{annala,margalit,radice,ruiz,shibata,most}. 
The stringent condition has forced to discard many exotic states of nature which were inferred to be at the center of the star. However, there is still a possibility that the interior of the star can have a deconfined state of matter.
With more GW detectors coming up shortly, our ability to localize and examine the sources will increase further. Both
the ground-based (like LIGO, Virgo, GEO600, TAMA300, Einstein Telescope) and space-based GW detectors (LISA) will have more efficiency with more detection 
capability (lesser amplitude GW).
Also, the detectors would cover a much broader range of frequency range which will help in the detection of more GW sources. So now we are in the age of long-promised GW and multimessenger astronomy. This will help us in tackling problems which are cross-cutting and 
multidisciplinary.

Neutron stars are one of the most exciting subjects of this GW and multimessenger astronomy as had been proved by BNSM GW170817 \cite{abbott1}.
The detection of the BNSM has proved beyond doubt that NSM is one of the best sources of GW. Even rotating NS which have time-varying quadrupole moment
can radiate GW and can be one of the sources \cite{ferrari,Zimmerman}. Fast rotating protoneutron stars can develop Chandrasekhar-Friedman-Schutz type
of instability and can generate enough energy to emit GW which can be detected \cite{chandrasekhar,friedman}. Non-radial oscillation can also be a possible
mechanism for NS to radiate GWs \cite{thorne}. Another event in NS that have enough opportunity to develop GW is the phase transition (PT) in NS \cite{prasad}.
All these potential GW detections can open a new window to probe NS interiors and consequently the matter at such nuclear densities.
The detection and interpretation of GW170817 have already provided the horizon of maximum mass, the radius and the tidal deformations.

The detection of GW from PT in NS is particularly interesting, for it being the signature of PT happening at extreme densities.
If such a confined deconfined PT does happen, it will prove the fact that exotic phases of matter do exist
at the interior of the stars. We then can have two families of compact stars coexisting. The existence of compact stellar objects with deconfined quark matter was predicted long
ago \cite{Itoh, Bodmer, Witten}. Such stars usually originate from NS by the conversion of hadronic matter (HM/NM)) to quark matter (QM). Such a PT can result soon after the birth in a supernovae \cite{bombaci,drago,mintz,gulminelli} or could happen in cold NS by accreting matter from its companion. The mass accretion would spin up the star (to millisecond pulsars), and there would be a large family of stars with high mass and rotation speed. These are the best candidates to suffer PT at the center of the star which suffers small fluctuations at the center.
The PT would likely be first order PT, where the nuclear matter is converted to quark matter (QM). As the density at the center of the star is maximum, the PT would in all probable to start from the center of the star \cite{prasad}. A combustion front or a shock wave is likely to propagate
from the center to the surface of the star. This front converts the NM to QM. The combustion front can spread throughout the star and reach the crust, thereby breaking it and ejecting matter or the front can lose steam inside the star and stop somewhere in between. 
The former case would result in a strange stable star (SS) whereas the latter would result in a hybrid star (HS). Such exotic stars are more compact than the original NS. Such PT would result in a release in the energy of the order of $10^{52}-10^{53}$ ergs \cite{bombaci1,berezhiani,drago,sahu}. Some amount of energy will be released in the form of neutrino emission. However, a significant amount of it can go into emitting gravitational waves \cite{lin,abdikamalov}.

In literature calculattion of gravitational wave emission from phase transition has been obtained via different 
approaches. In one of the approach, the oscillation modes excited by PT is said to result in GW. The gravitational waves produced due to any given mode is given by,
\begin{equation}
 h(t)=h_{o} e^{-\frac{t}{\tau}}  sin(\omega_{0} t)
\end{equation} 
where $h_{0}$ is amplitude given by
\begin{equation}
h_{0} = \frac{4}{\omega_{0} D} \sqrt{\frac{E_{g}}{\tau}}
\end{equation} 
where $E_{g}$ is the energy available to give a oscillation mode, $\omega_{0}$ is frequency of oscillation mode and  $\tau$ is damping time scale corresponding to a give mode. 
The energy released in PT is estimated using the difference in binding energy of NS and HS/QS, given by
\begin{equation}
B.E. (r) = M_{g}(r) - M_{B}(r).
\end{equation} 
During phase transition, the energy released is majorly in the form of gamma-ray burst and gravitational waves. Also, there will be other dissipation mechanisms, hence in this approach the energy budget available to gravitational waves is uncertain. In other approach implemented by Lin et al. \cite{lin} and 
Admikamalov et al. \cite{abdikamalov}, it is considered that quark matter content when appears inside an NS, due to quark matter EoS having less pressure than nuclear matter the star undergoes a micro-collapse. This results in quadrupole moment change, and the gravitational wave is generated.
The calculation of the GW by Lin et al. \cite{lin} treats the PT to occur instantaneously. The collapse would result in stellar pulsation and using Newtonian gravity they obtained the waveforms of the emitted GW from such collapsar models and found that the strain of the GW can of the order of $10^{-23}$. This calculation was further carried forward by Abdikamalov et al. \cite{abdikamalov}, where they used GR 
calculation and more realistic EoS models. The PT was assumed to be of finite timescale, but the GW calculation involved were related to the stellar pulsation of different modes.

This approach deals mainly with the aftermath of PT. The gravitational waves resulting from the dynamical process of PT has not been investigated. In our recent work \cite{prasad} (let call it Phase transition in neutron stars (PTNS)), 
we have simulated the dynamical PT process using a conversion font (shock front) which propagates from the center to the surface and converts quark matter to nuclear matter.
In PTNS we have studied the actual dynamical evolution of the combustion front in NS. The conversion of NM to QM takes place at this combustion front, and as the front moves outward to the outer surface of the star
the shock intensity decreases and the combustion stops inside the star, however, the velocity of the shock propagation is close to that of light. This indicates a rapid PT, with the timescale, of the order of tens of microseconds. If such a PT happens inside a rapidly rotating neutron star having an axisymmetric shape, the quadrupole moment of the star changes at this small time scale. This can result in a sufficiently strong GW signal, which can be detected by either improved earth based GW interferometers or by space-based GW detectors. In this article, we report the characteristics and the template of such GW signals that are likely to be produced in such PT of NS.

In section II we first describe briefly our equation of state (EoS) which we have employed to characterize the QM and HM. Section III is devoted to the description of the numerical model of the star and the combustion of NS to QS. The GW results are given in section IV and finally, in section V we discuss our findings and conclude from them.

\section{Equation of State}

NS matter at the inner core is very dense, and they interact via the strong interaction. In principle the degree of freedom at such densities should be quarks; however, the quarks appear at the core after some fluctuation in the star (after PT). Therefore, to begin with, the degree of freedom is consequently mainly neutron, proton, electron and some other baryons and leptons in a small fraction. The carriers of the nuclear force are assumed to be sigma, omega, and rho. In this calculation we choose PLZ \cite{reinhard} 
parameter setting to describe the NM of the star leaving the crust. The EoS is consistent with the recent astrophysical and nuclear constraint and can generate stars more massive than two solar mass. The equatorial radius of the star of mass $1.5$ solar mass is about $15$ Km.

After the PT the degree of freedom of the innermost core of the star becomes quarks. The QM is described by the MIT bag model having quark interaction \cite{chodos}. The QM is composed of only up and down quark. The shock propagation deconfines the hadrons to quarks (2 flavor quark matter). This happens very fast, and this 2 flavor (2-f) matter is metastable. It settles into final stable strange matter (3-f) (with strange quark appearing via weak interaction) at weak interaction time scale. However, this process is much slower than the former deconfinement of quarks \cite{abhijit}. Therefore, we can treat the two process separately. In this problem, we are dealing with the first process. For the quark matter, the bag value is chosen to be $B^{1/4}=140$ MeV and the quark coupling to be $a_4=0.5$.

\begin{figure}
%\vskip 0.2in
\includegraphics[width = 3.14in,height=2.0in]{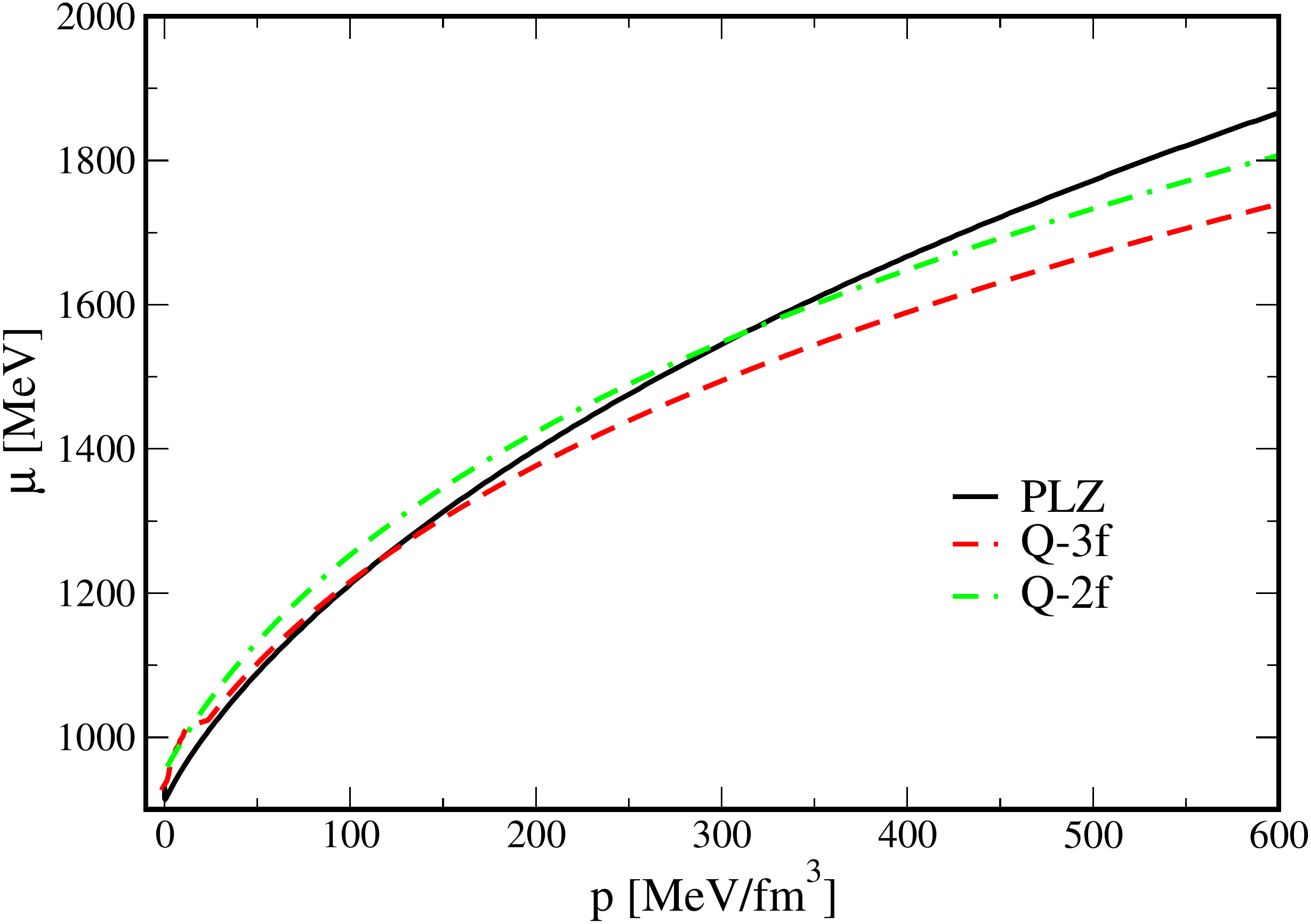} 
\caption{(Color online) The chemical potential is shown as a function of pressure for PLZ, 2-f QM and 3-f QM.
The $B^{1/4}=140$ MeV 2-f curve is the metastable QM, and the 3-f curve is stable QM.
The points where the PLZ and 2-f QM curves coincide depicts the equilibrium PT from hadronic to quark matter.}
\label{fig-2f}
\end{figure}

Fig \ref{fig-2f} shows that the 3-f QM (Q-140) is absolutely stable than the HM (PLZ) beyond $p=110$ Mev/fm$^3$. However, the metastable 2-f starts at $p=340$ Mev/fm$^3$ and beyond those densities 2-f QM is stable. In our scenario where the PT happens via a 2-step process, the NM first has to convert to 2-f matter and then to 3-f matter. If the NM is stable than 3-f QM, it will not convert to QM by such a process. The PLZ and 2-f curve (Q-140) cuts at $p=110$  and below those pressure, the 2f QM is metastable and finally has to go to 3-f QM for absolute stability. This, tells us that for a PT taking place from PLZ NS to a quark star (QS) (with Q-140), 
the final QS is likely to be a hybrid star (HS). Therefore, we only have QM beyond pressure $p=110$ Mev/fm$^3$. In our star, this happens at a distance of $6 km$, and we have the PT or combustion from NM to QM till a distance of $6 km$ and beyond that radial distance we have shock propagation.
For the quark matter, the bag value is chosen to be $B^{1/4}=140$ MeV and the quark coupling to be $a_4=0.5$.

\section{Hydrodynamic simulation of the combustion of NM to QM}

This work aims to calculate the strain and power of the GW which would be emitted from such a PT. Shock-induced PT in spherically symmetric neutron star dosen't lead to GW emission, hence in the present study the neutron star is taken to be axisymmeteric. To solve for an axisymmetric star using the given EoS we employ the rotating neutron star (RNS) code\cite{komatsu,Stergioulas}. The structure of the star is described by the  Cook-Shapiro-Teukolsky (CST) metric \cite{cook}. The code can solve for rotating NS using polytropic or tabulated EoS. The code calculates the metric functions of the metric (they are a function of $r$ and $\theta$). It takes the central density and rotational velocities as an input and give back the metric function in terms of $r$ and $\theta$. It also gives the density, pressure as a function of the dependent variables.  The total star mass, its equatorial and polar radius, its moment of inertia and its eccentricity are also obtained.  
Using PLZ EoS neutron star strucutre and properties is obtained for different values of central density keeping the rotational frequency $\omega = 0.3 \times 10^{-4}$ Hz fixed, these different NS models obtained are listed in table 1 and would be used in our calculation of the combustion and also for the GW.

The hydrodynamic equation is solved using the GR1D code \cite{oconnor,font}. The description of the code can be found in our previous paper, and we only discuss them briefly  in this present article \cite{prasad}.The code uses the method of lines for time integration \cite{hyman}. The spatial discretization is done by finite volume approach \cite{romero,font}. All the primitive variables are defined at the cell center and are interpolated at cell interfaces. Piecewise-parabolic method (PPM \cite{Colella}) is for interpolation of smoothing of fluxes. From the primitive variables the conserved variables are calculated (density, velocity, and energy). The physical fluxes are calculated by 
HLLE Reimann solver \cite{font}. Once the conserved variables are calculated the EoS is used to calculate the pressure. So, we have all the thermodynamic variables evolving both spatially and temporally.

The solution of the RNS code is used to set the initial configuration at some particular $\theta$. This gives the star profile (density, pressure as a function of radial distance for a specific $\theta$. For our specific problem, we initially give a small density fluctuation near the center of the star (at about $0.5 km$). At an initial time $t=0$, the velocity of both left and right states are zero. As the shock propagates the position of the shock discontinuity concurs with the location of which the speed is maximum. The position of the shock discontinuity can be traced for all time using this method. We ensure that in behind (QM EoS) and front (HM EoS) of the shock the EoS is different, ensuring a simulated PT.

In PTNS we have simulated the shock-induced phase transition using one-dimensional general relativistic hydrodynamics code (GR1D). 
To obtain a PT, we need a significant density fluctuation at the center of the star. The change can be caused due to sudden spin up due to mass accretion. We heuristically assume a density discontinuity at a distance of $0.5$ km from the center of the star. This gives rise to a shock wave. With time the shock wave propagates outward and thereby combusting the NM to QM. The shock wave is evolved according to the solution of the hydrodynamic equation. This is done by the GR1D code. We do a 2-dimensional evolution of the shock profile (in $r$ and $\theta$) solving the GR1D code for different $\theta$ obtained from the RNS code. The initial discontinuity is given for the density, and the pressure discontinuity is obtained from the polytrope. The initial matter velocities at either side of the shock are kept to be zero. We see that with time the discontinuity proceeds towards the periphery of the star from the high-density to the low-density region and its strength gets reduced. To start with, we carry out our study using the PLZ-M1 model having central density $2$ times the nuclear density and $\omega = 0.3 \times 10^{-4}$ Hz, which yields a star of mass $1.5 M_{\odot}$ and radius of $15$ km.

We proceed with following steps to ensure that system in hydrodynamics code is a neutron star in 2-dimension and mimics realistic scenario whereas the conversion font propagates the hadronic matter region which it surpasses becomes 2-flavor quark matter. We use RNS code to obtain the profile of neutron star using hadronic matter EoS; it gives us a profile for an axisymmetric neutron star. The $ 0 < \theta < \pi/2 $ is split into 65 values. For a given $\theta$, we obtain $\rho (r, \theta_{fixed})$. These 65 profiles have different values of $r_{max}$ the boundary of the star.  The hydrodynamics equations require initial conditions which are the value of density, pressure, and velocity at each point in the system at $t=0$ and the boundary of the star. We evolve this $65$ profiles(each corresponding to a $\theta$) one by one. The overall output of hydrodynamics simulation comes out to be density $\rho (r, \theta_{fixed})$, pressure $p (r, \theta_{fixed})$ and velocity $v (r, \theta_{fixed})$ at each point in the system at any time $t$ along a given direction. By combining all these $65$ profiles along $65$ directions, we get the complete information of star. In fig 2., we show the density variation of the star (before the PT) along the radial distance for PLZ-M1 case. Fig 2 and table 1 are in agreement and shows that the rotation in the star deforms the star. The star is oblate spheroid, with the equatorial radius is elongated than the polar radius. The density (and pressure) fall smoothly from the center to the surface for an unshocked star.

\begin{figure} 
\includegraphics[width = 3.5in]{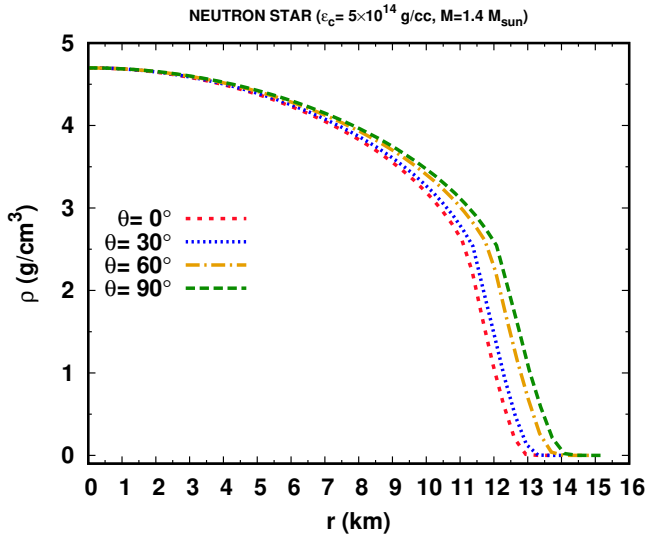}
\caption{The density variation along the star radial distance is shown in the figure. The variation is shown for four different angle ($\theta=90^{\circ}$ (equator), 
$\theta=60^{\circ}$, $\theta=30^{\circ}$ and $\theta=0^{\circ}$ (pole). The eccentricity of the star is $0.425$.}
\label{star-prof}
\end{figure}

\begin{figure*}
%\vskip 0.2in
\centering
\includegraphics[width = 3.5in,height=2.8in]{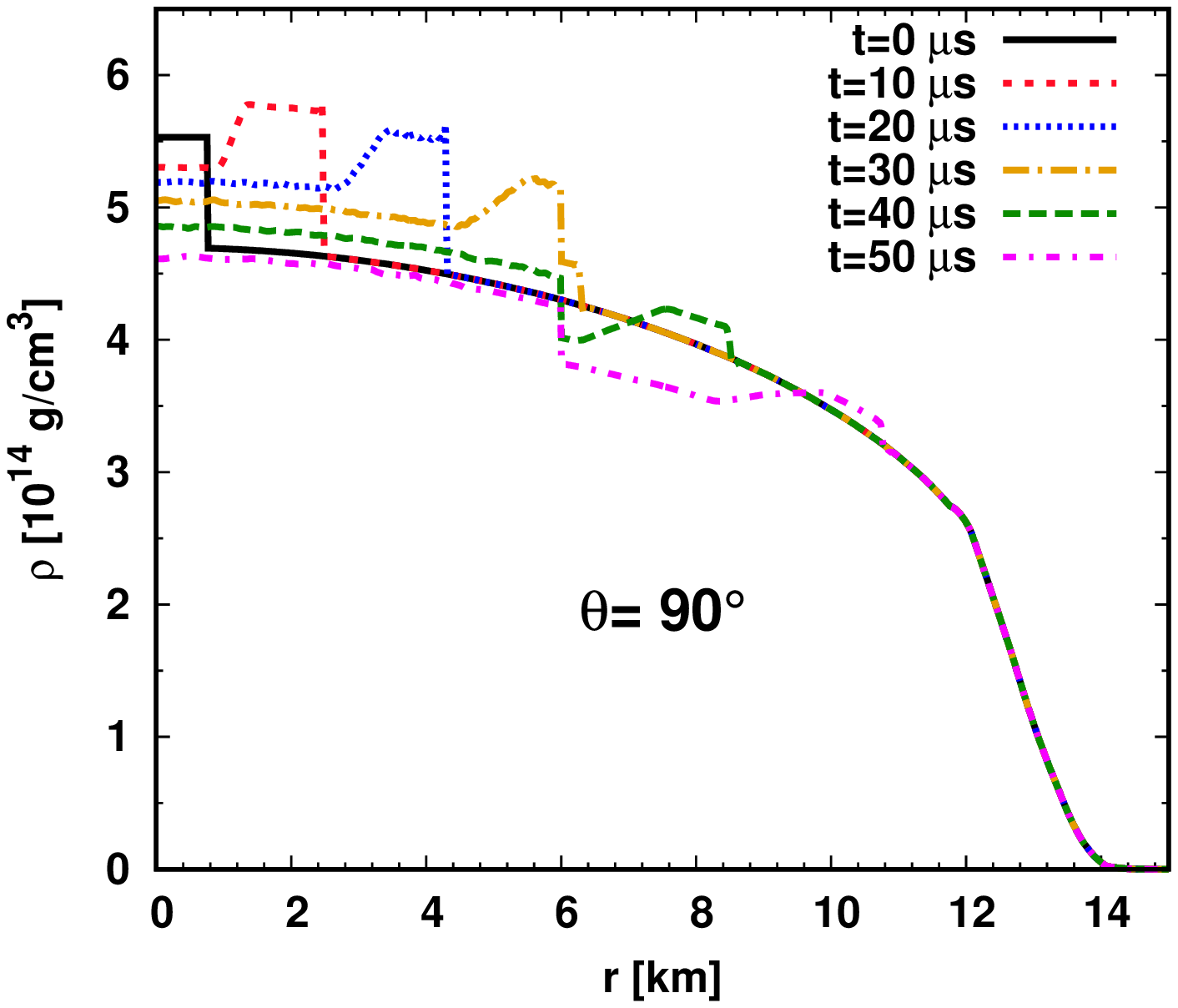}
%\hskip .4 cm
\includegraphics[width = 3.5in,height=2.8in]{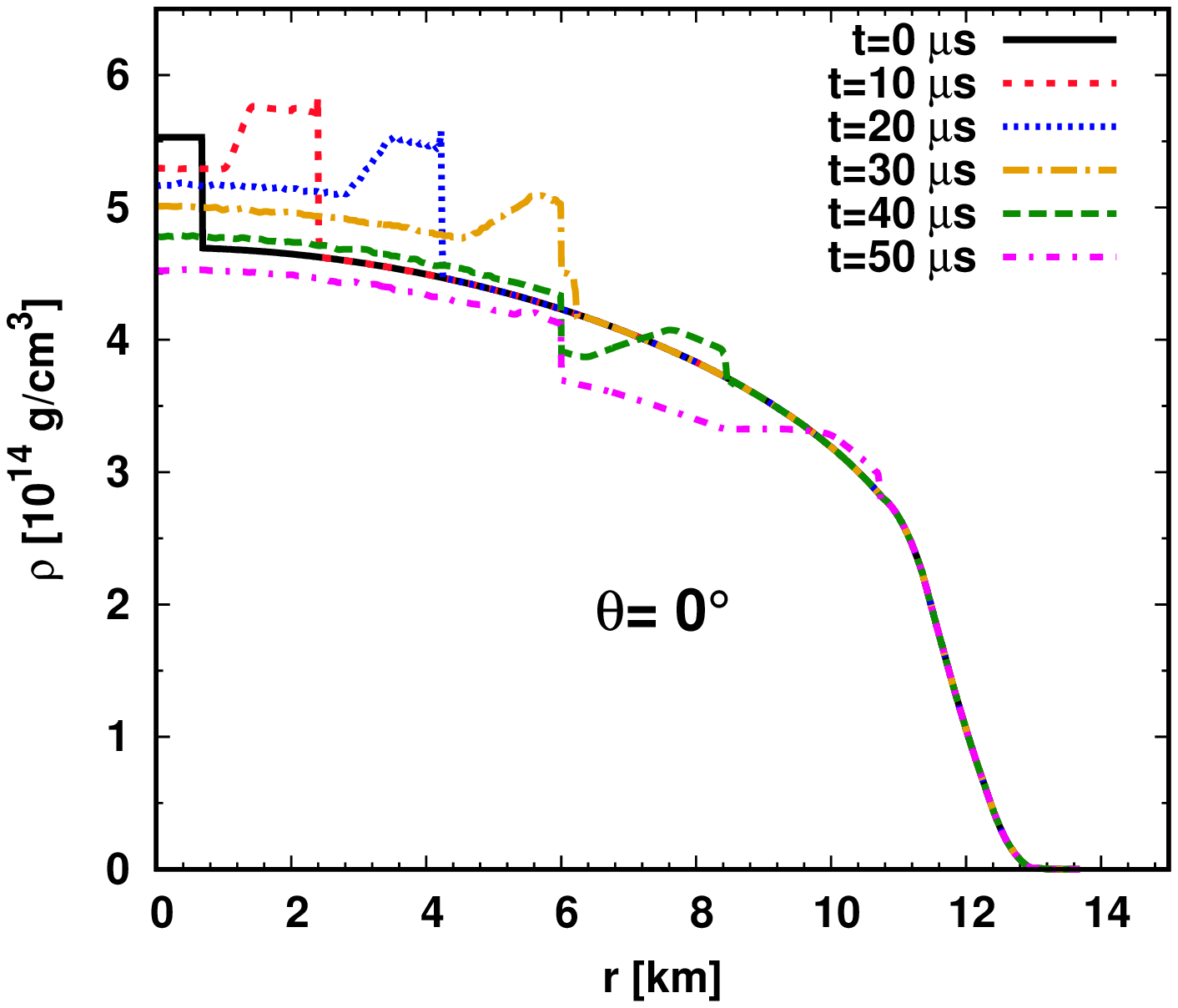}

\hspace{0.5cm} \scriptsize{(a)} \hspace{7.8cm} \scriptsize{(b)} 
\caption{(Color online) The evolution of density as a function of radial distance of the star is shown. After an initial density fluctuation we have shown the density
evolution due to a shock propagation for four instances of time. The PT occurs till $7$ km and after that simple shock propagation is observed. The plot is done for the propagation 
along a) equatorial and b) polar direction.}
\label{rho}
\end{figure*}

\begin{figure*}
%\vskip 0.2in
\centering
\includegraphics[width =3.5in,height=2.8in]{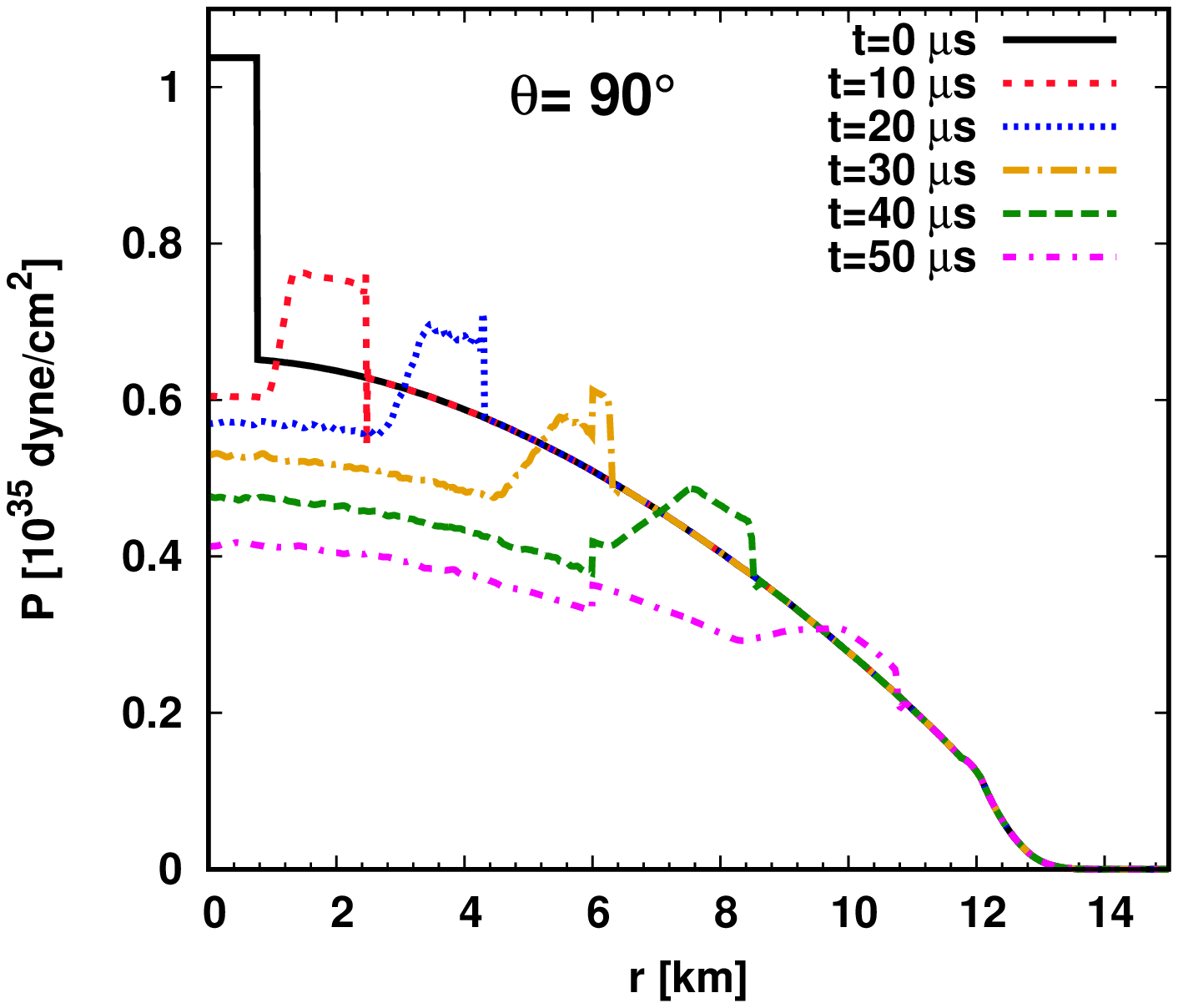}
%\hskip .4 cm
\includegraphics[width =3.5in,height=2.8in]{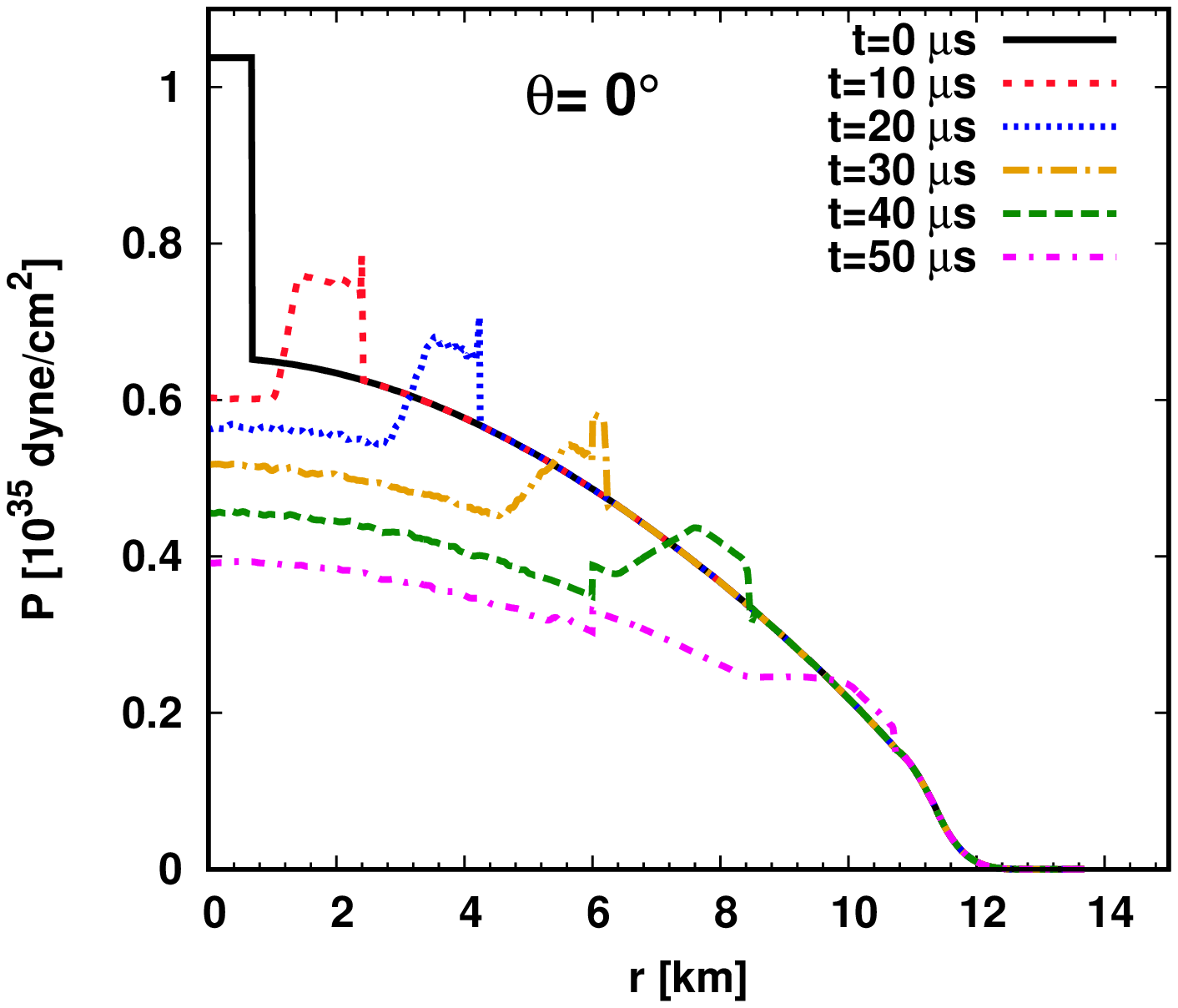}

\hspace{0.5cm} \scriptsize{(a)} \hspace{7.8cm} \scriptsize{(b)} 
\caption{(Color online) The density fluctuation results also in the discontinuity of pressure. The evolution of pressure as a function of the radial distance of the star is plotted. After an initial density fluctuation, we have shown the evolution of
pressure due to a shock propagation for four instances of time. The PT occurs till $6$ km and after that simple shock propagation is observed. The plot is done for the propagation 
along a) equatorial and b) polar direction.}
\label{prs}
\end{figure*}

\begin{figure*}
%\vskip 0.2in
\centering
\includegraphics[width = 3.5in,height=2.8in]{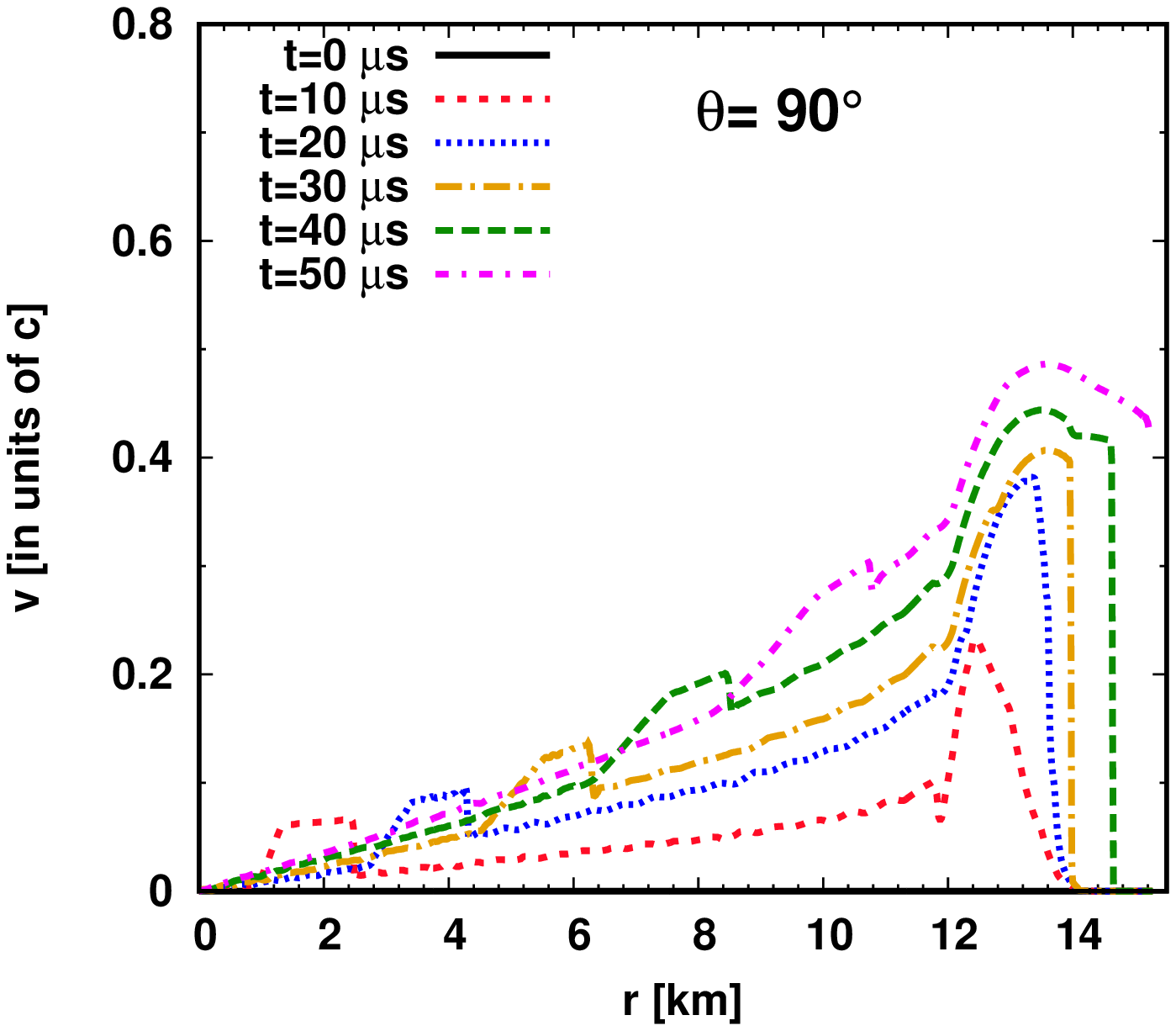}
%\hskip .4 cm
\includegraphics[width = 3.5in,height=2.8in]{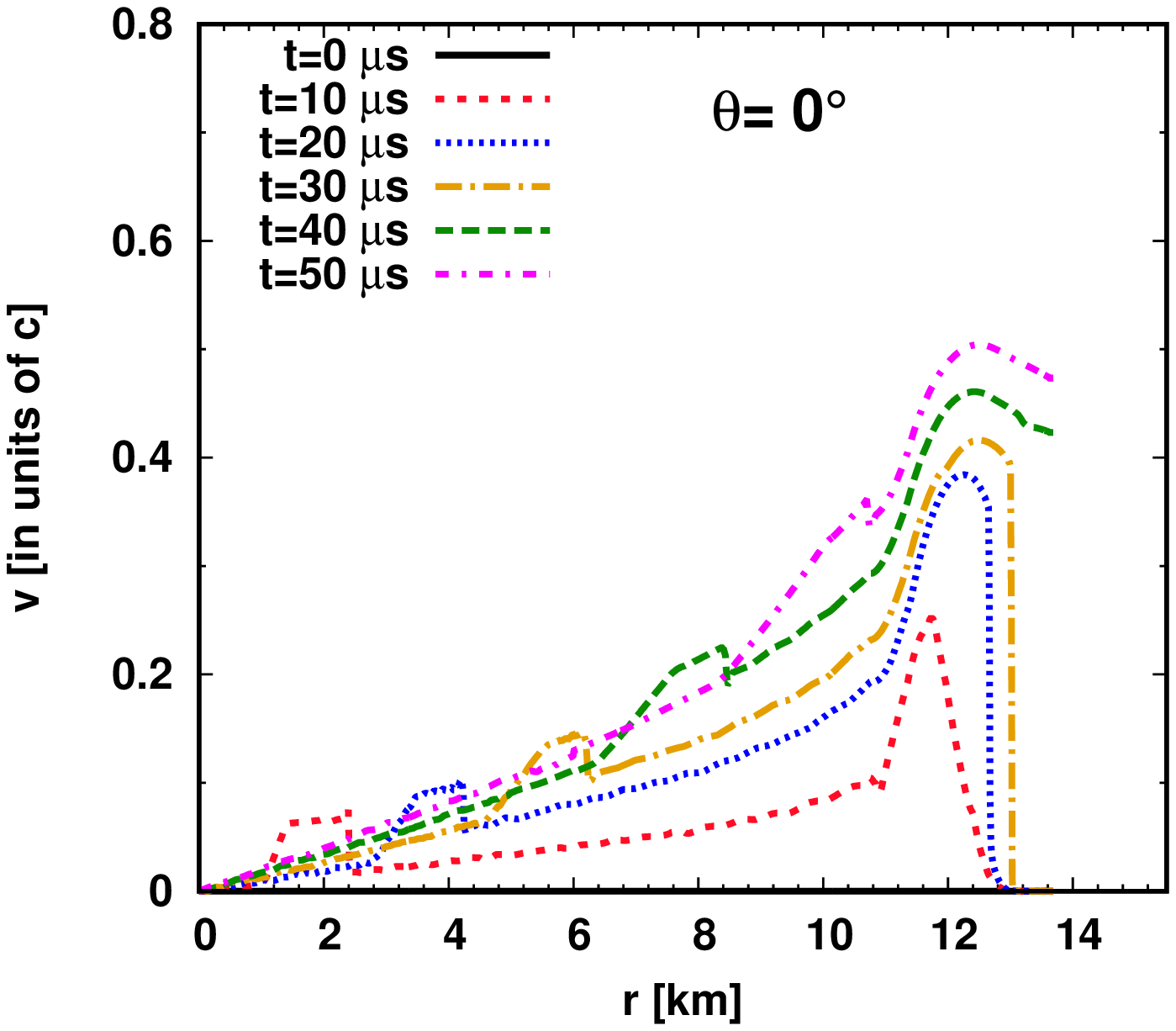}

\hspace{0.5cm} \scriptsize{(a)} \hspace{7.8cm} \scriptsize{(b)} 
\caption{(Color online) The matter velocity at initial time is zero (no combustion wave). As the combustion starts, the matter velocity takes non zero values. We have shown the 
value of matter velocity as a function of radial distance. The plot is done for the matter velocity  
along a) equatorial and b) polar direction. }
\label{vel}
\end{figure*}

\begin{figure} 
\includegraphics[width=3.5in, height=2.8in]{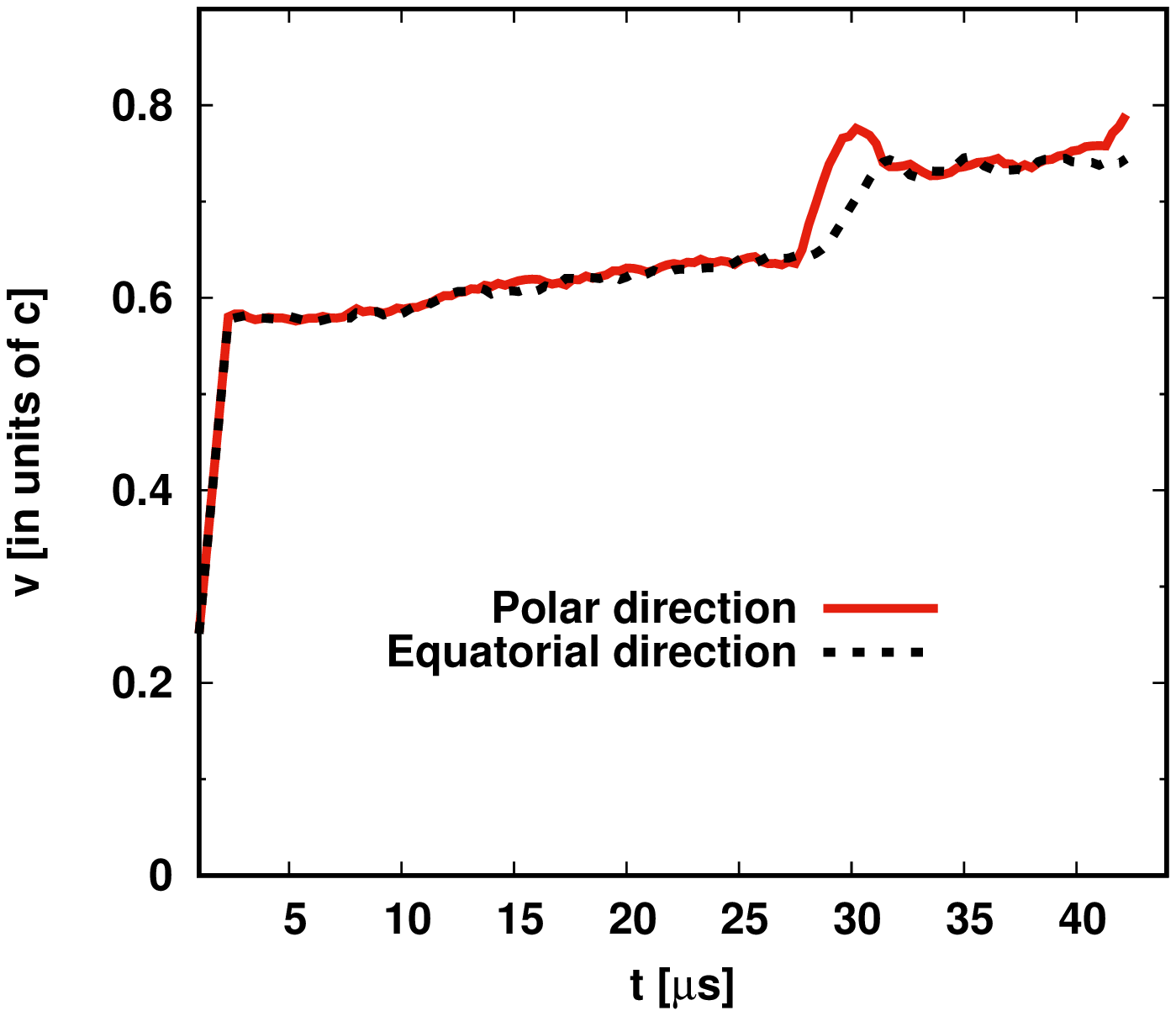}
\caption{The velocity of the combustion front and the shock wave is shown as a function of time. Till $25 \mu s$ the combustion wave propagates from the center to a distance of $6$ km.
After $6$ km simple shock wave propagates to the surface of the star. The transition from combustion wave to shock wave is seen as a sharp discontinuity in the front velocity. The front 
velocity is also plotted for the equatorial and polar direction.}
\label{vel-f}
\end{figure}
As the shock wave propagates, it converts NM to QM, however after some distance as the density decreases the shock wave starts to lose its strength. Also if there is a formation of the mixed phase, it will dissolve this sharp discontinuity. The smoothing of the discontinuity will depend on how much time needed for the mixed phase region to grow, whereas our propagation of the shock is swift. It is quite possible that it can travel a considerable distance before it smooths out. Therefore, after propagating some distance in the star, the PT will stop and eventually the shock wave may die out. We should mention that we have not done the exact calculation on how long it would take for the sharp discontinuity to dissolve due to the formation of the mixed phase. Such a calculation although may be essential but involves much complexity. For simplicity, we do not consider such a case in our study. The lowest density of the quark matter EoS which we have used is about $1.8$ time the nuclear density, which corresponds to a distance of about $6$ km in the star. Therefore, after a length of $6$ km, the shock evolves without bringing about a PT. We, therefore, have an HS, a star with outer NM and a quark core. 

\begin{figure*}
\centering
\begin{minipage}[t]{7.2in}
{\includegraphics[height=2.95in, width=3.1in]{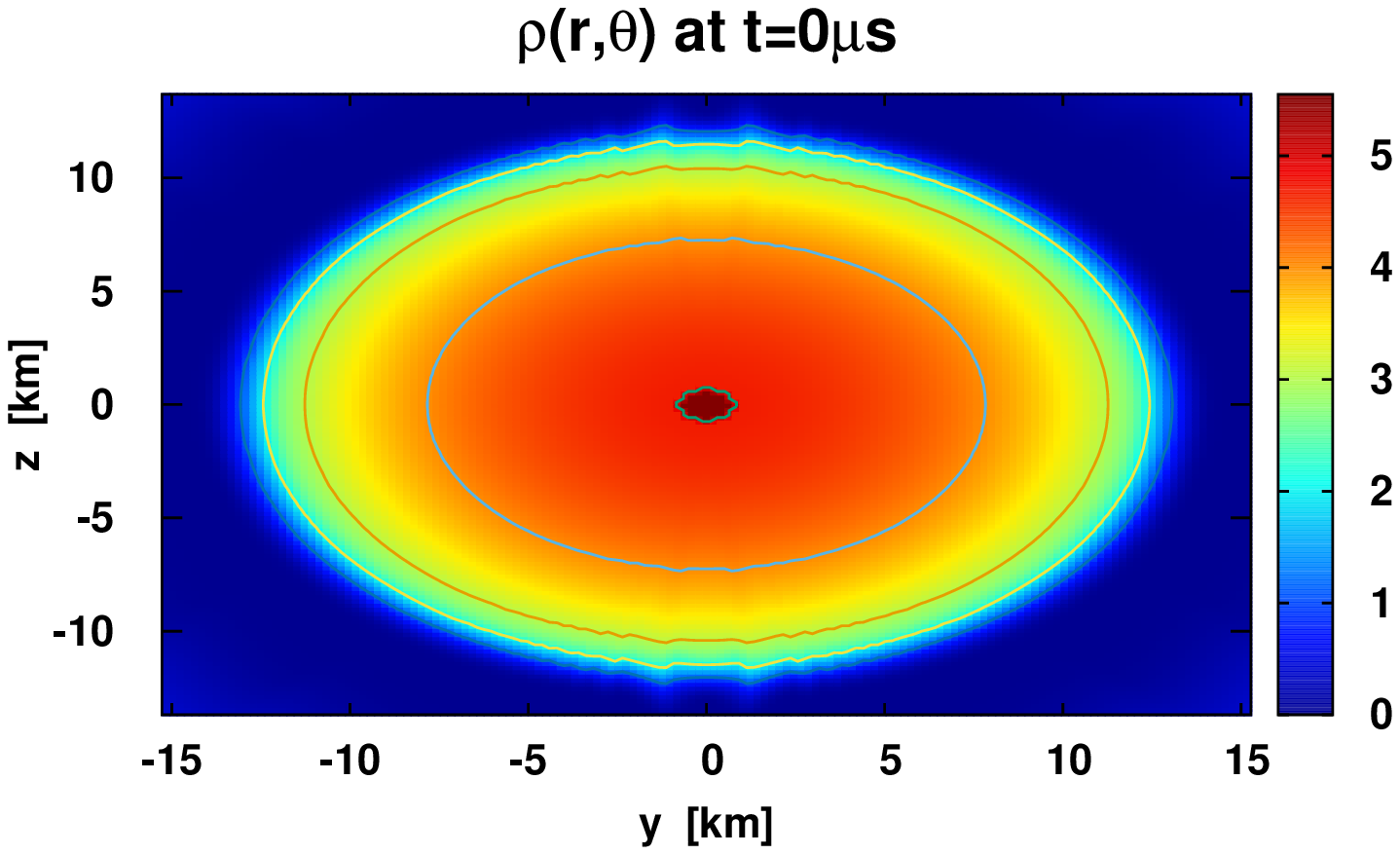}}\hspace{0.5cm}
{\includegraphics[height=2.95in, width=3.1in]{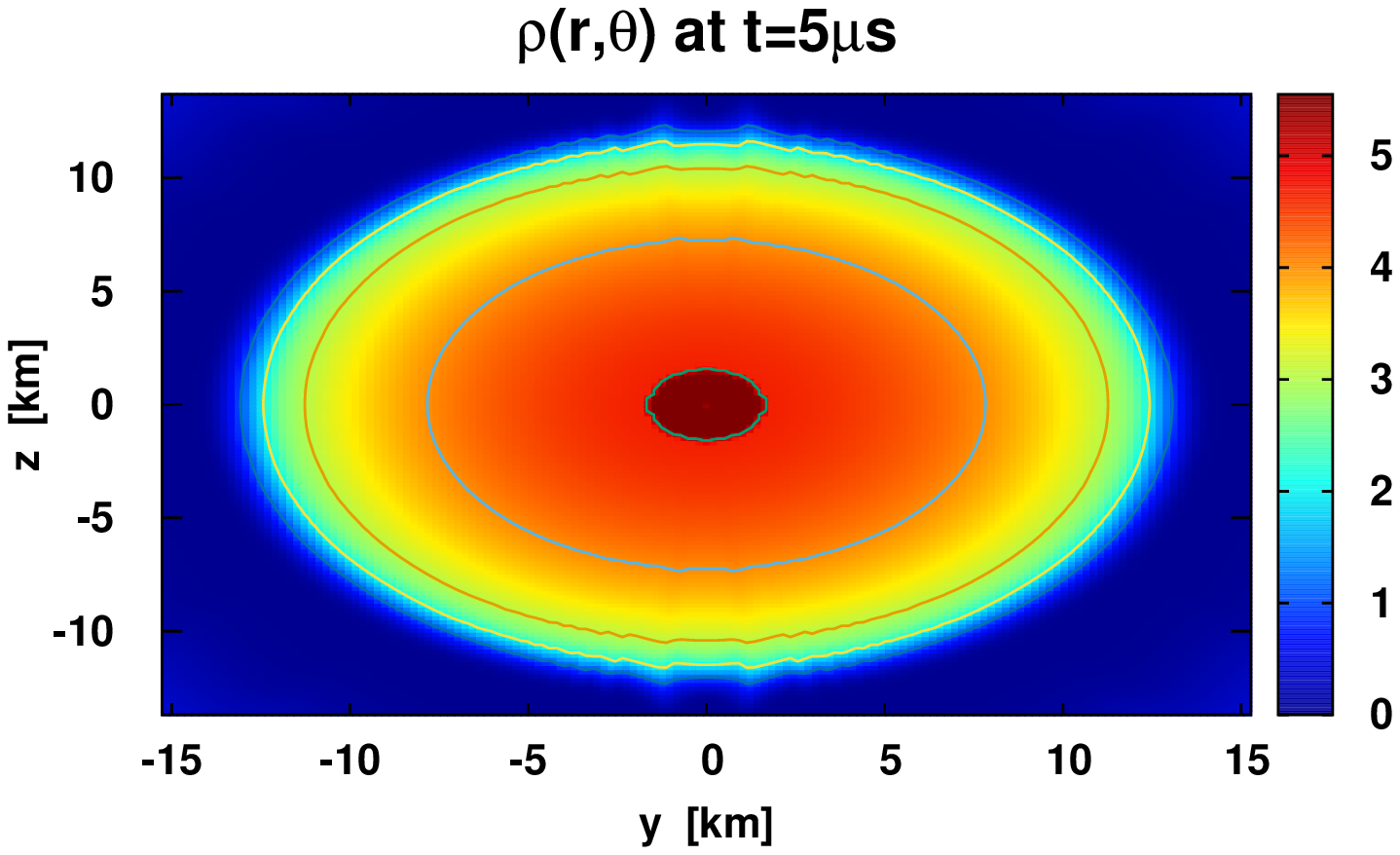}}%
\end{minipage}%
\vspace{0.2cm}
\begin{minipage}[b]{7.2in}
{\includegraphics[height=2.95in, width=3.1in]{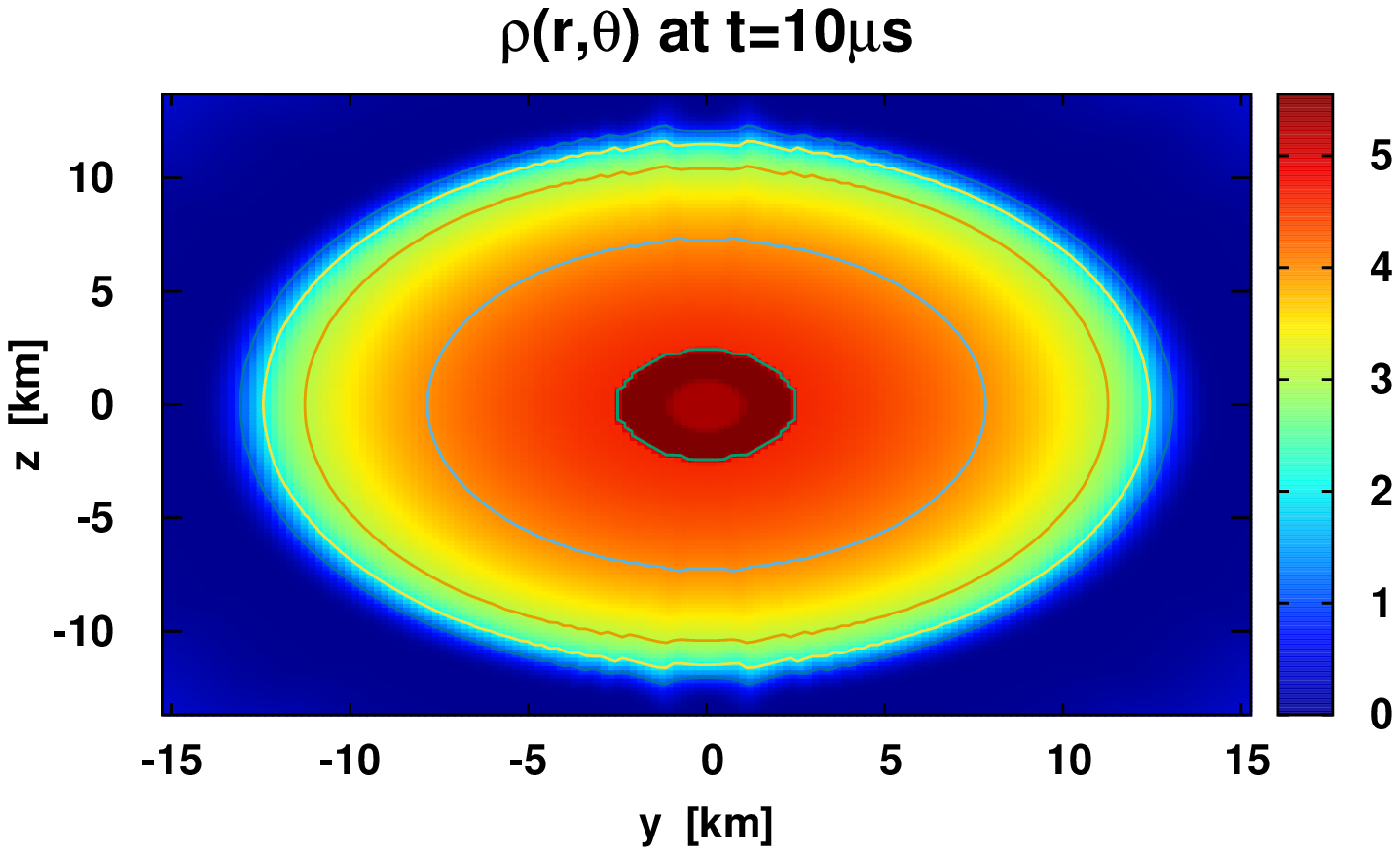}}\hspace{1cm}
{\includegraphics[height=2.95in, width=3.1in]{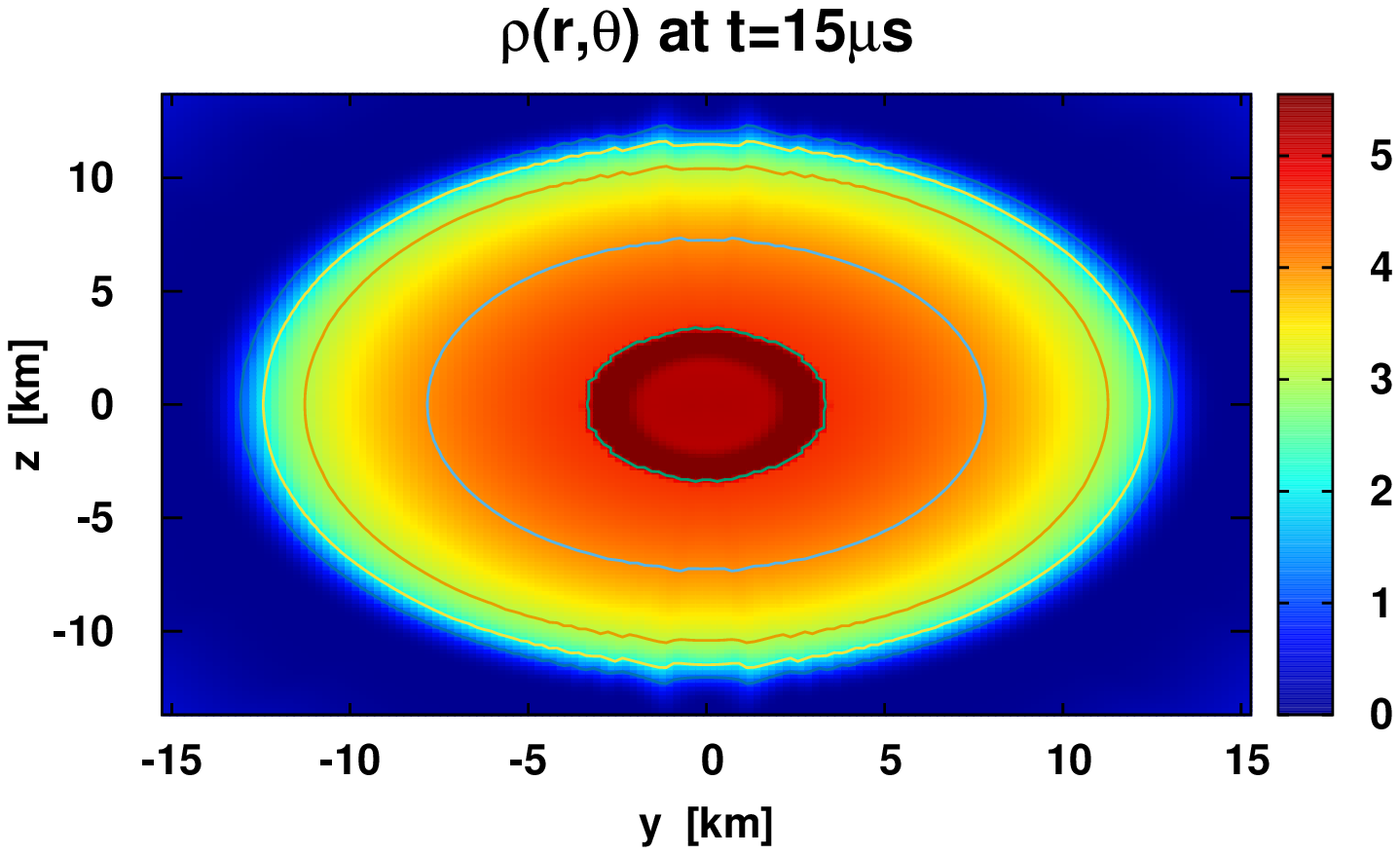}}
\end{minipage}%
\vspace{0.2cm}
\begin{minipage}[b]{7.2in}
{\includegraphics[height=2.95in, width=3.1in]{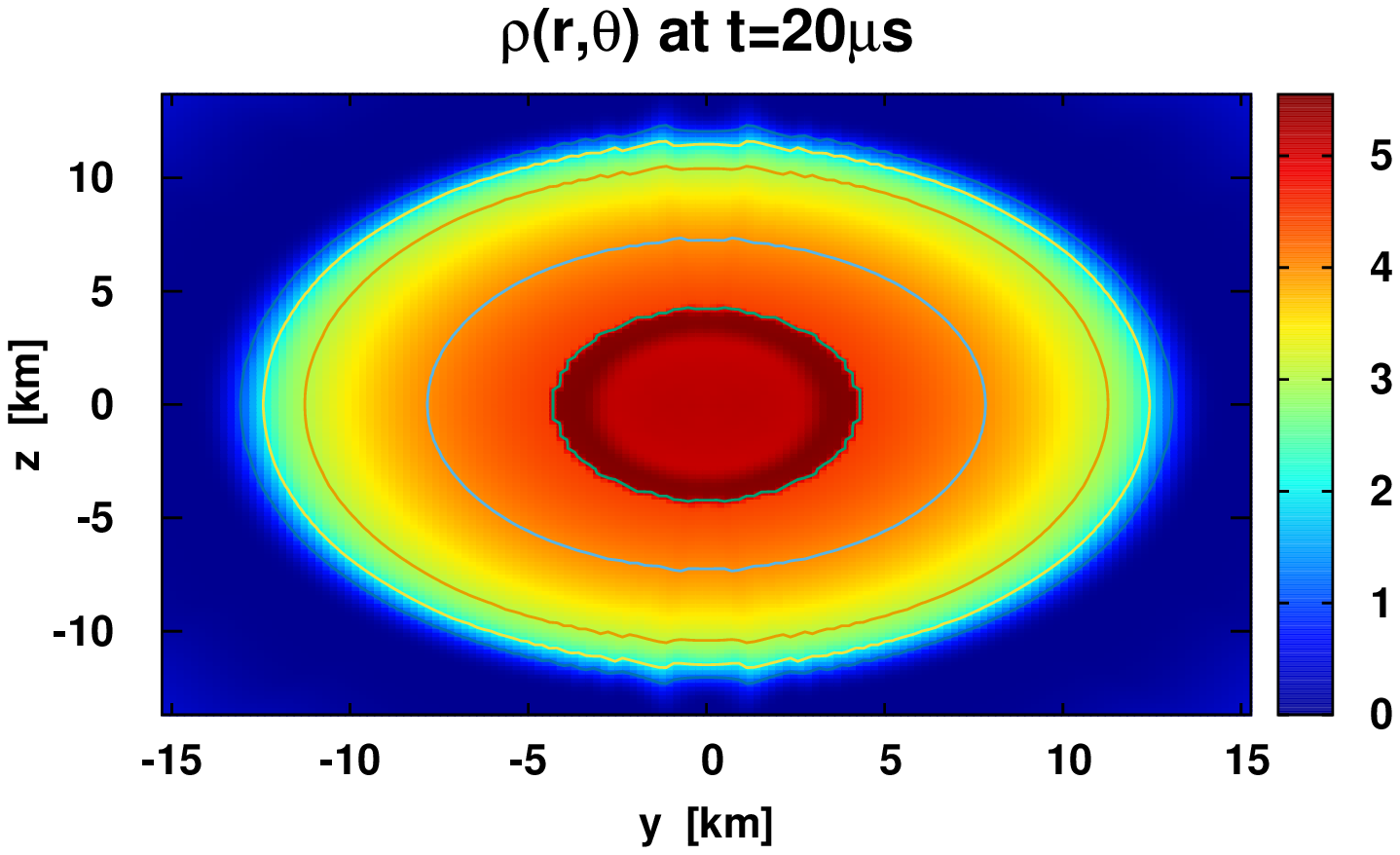}}\hspace{1cm}
{\includegraphics[height=2.95in, width=3.1in]{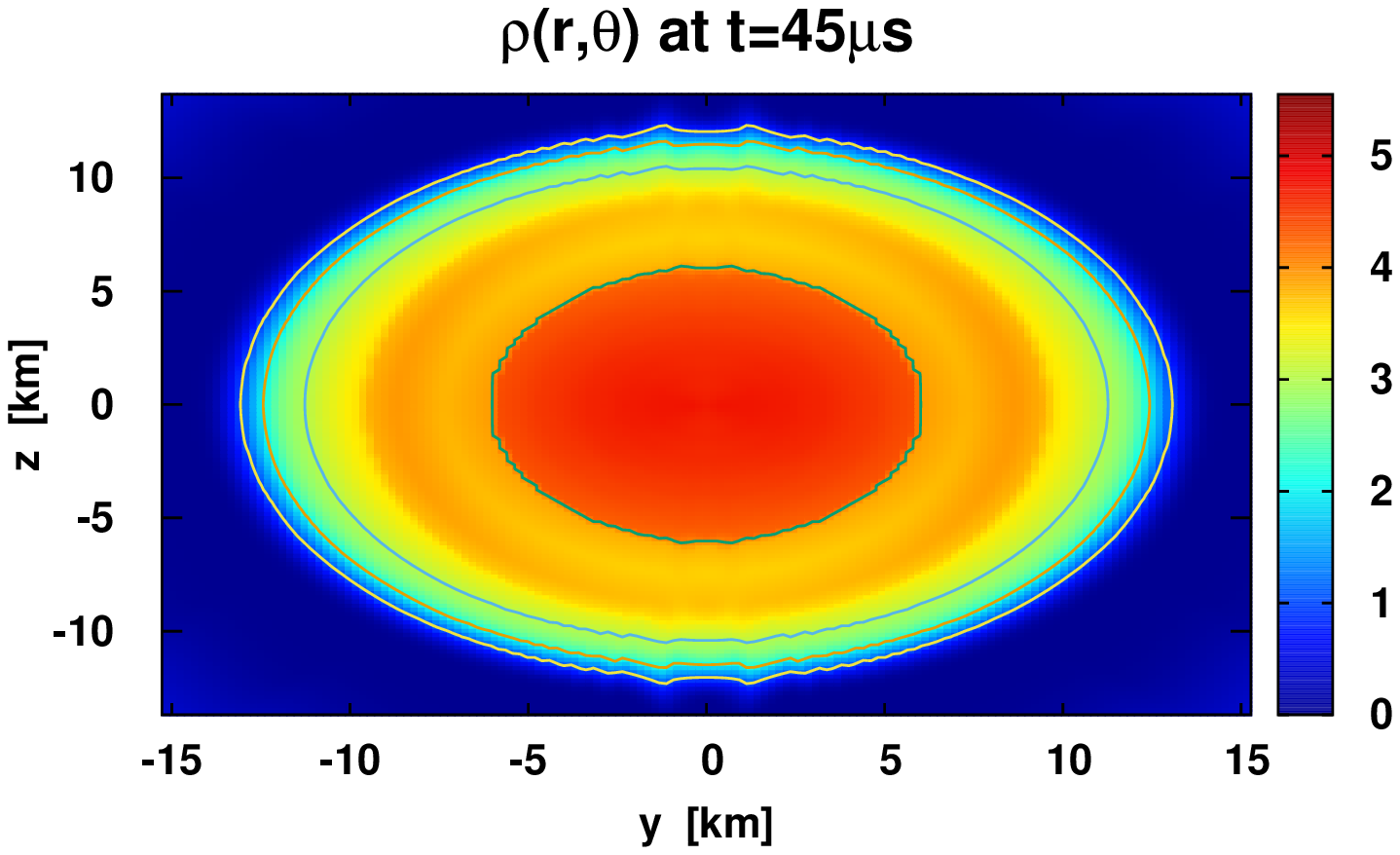}}
\end{minipage}%
 \vspace{0.2cm}
\caption
{%
PT in neutron star (PLZ-M1) for different time instances. The horizontal axis represents equatorial direction and vertical axis represents polar direction. The PT front as propagates outward it converts the region surpassed by it from nuclear matter to quark matter, which can be seen as increasing density in this plots.
\label{fig:caption}%
}%
\end{figure*}

Fig 3 shows the density evolution as a function of the radial distance of the star. We have demonstrated this evolution for the equatorial and polar direction. The hydrodynamic equations also solve for the spatial and temporal evolution. Therefore, we have shown $6$ time slice (at four different times including initial configuration) plots of the density and pressure evolution as a function of the radial distance.  
The EoS ensures that a discontinuity in the density results in a discontinuity in the pressure of the star. In fig 4 we plot the pressure as it evolves both spatially and temporally. The evolution of the pressure strictly follows the development of the density. The hydrodynamic equation ensures that as the combustion starts the matter velocities takes up non zero values. The matter velocities are shown in fig 5. As the frame moves outwards, from the frame of the front, the HM comes towards the front. At the front, it is converted to QM, and beyond the front, the QM goes away from the front.  As the star is a closed and a dense system the velocity of the matter peaks up at the surface. This may result in the breaking of the crust and ejection of matter to the star atmosphere. However, gravity may play a crucial role here. In this article we have not taken gravity into account; therefore we refrain from further comments. The density or the pressure plots gives the location of the combustion or the shock front at any given instant of time. Therefore, we can calculate the front velocity by differentiating the shock location with respect to time. In fig 6 we show the shock velocity along the polar and equatorial direction. More or less the nature of the front velocity remains the same. We assume that the combustion wave starts from $0.5$ km and continues till $6$ km. Beyond that only shock wave propagates. Therefore at $6$ km there is a transition from combustion to shock wave. This is seen clearly as a sharp discontinuity of the front velocity. The combustion wave takes around 30 $\mu s$ to travel to $6$ km in the star. This sharp discontinuity has a significant contribution to the GW signal which can be seen later. The 2-dimensional evolution of PT front is shown in fig 7, where for $\phi= \pi/2$ the $\rho=(r,\theta)=\rho(z,y)$ is represented as heatmap for different time instances.

\section{Phase transition as source of quadrupole moment change}
%\begin{widetext}
\begin{table*}
\center
\setlength{\tabcolsep}{3pt}
\renewcommand{\arraystretch}{1.5}
\caption{Stellar models}
\begin{footnotesize}
\begin{tabular}{ccccccccc}
\hline 
\hline
Model & \multicolumn{6}{c}{Star Properties}& \multicolumn{2}{c}{Rise $\approx 1.18 \rho_{c}$}\\ 
\cline{2-7} \cline{8-9} 
 &$\rho_{c}$&$\omega$&$M$&$r_{e}/r_{p}$&$T/W$&I&$|h|$&$f_{peaks}$\\
  & ($10^{14}$ g/cc)&($10^{4}$ Hz) &($M_{sun}$) &  &($10^{-2}$)& $10^{45}$g cm$^{2}$ & &kHz\\   
\hline
PLZ-M1 &5.0& 0.3& 1.50& 1.28871& 2.73182& 2.22& $10^{-22}$& 23.80, 95.23, 166.66, 261.90\\
PLZ-M2 &5.6& 0.3& 1.82& 1.22754 &2.44001& 2.96& $10^{-22}$& 23.80, 95.23, 166.66, 238.09\\
PLZ-M3 &6.0& 0.3& 1.99& 1.19347 &2.29033& 3.39& $10^{-22}$ & 23.80, 71.42, 142.85, 285.714\\
PLZ-M4 &6.6& 0.3& 2.2& 1.14929& 2.10966& 3.92 &$10^{-22}$& 23.80, 71.42, 119.04, 190.47, 238.09  \\
\hline
\hline
\end{tabular}
\end{footnotesize}
\end{table*}
%\end{widetext}

Recent gravitational wave observation of GW170817 \cite{abbott1} of a neutron star merger has shed new light not only in the field of GW astronomy but also to multimessenger astronomy. The star merger has revealed itself in the form of GW, neutrino production and also in other various spectra of electromagnetic range. With the rejuvenation of GW astronomy and the GW detection, several new gravitational wave detectors are coming up shortly like TAMA300 of Japan, a space-based telescope LISA, Einstein Telescope along with the presently working LIGO, Virgo and GEO600. This detectors and telescopes are much improved and are likely to detect small amplitude GW and with frequency ranging upto KHz. The PT scenario is also a very credible source of generation of GW and the combustion time of the star hints at a very definite template of such a scenario. The small time scale of NM to 2-f QM conversion is quite different from any other astrophysical timescale and therefore if such a process can generate GW it is very likely that the signal of such a process would also be very different.

In this section, we calculate the prospect of GW signal due to the combustion of NM to 2-f QM. We have done a relativistic hydrodynamic simulation of the combustion process; however, gravity is not taken into account. The gravity is likely to influence the process, but it is expected that it will not have any considerable effect on the combustion process due to fast burning. The gravity is more likely to have a significant impact on the 2nd step process (2-f QM to 3-F QM) where the star tries to settle down and suffers a gravitational collapse.
Such process has been readily studied in the literature \cite{dimmelheimer, lin,abdikamalov} and it suggests that the collapse would last few milliseconds and the GW emission from such a process would have an amplitude of the order of $10^{-23}$. The time frame mentioned in those papers is of the order of 2-f to 3-f (weak interaction) time scale. The possible problem with such a GW signal is that it hints at any type of collapse dynamics. It may come from a PT but even can generate from some reshuffling of the density profile of the star. However, a signal of NM to 2-f QM conversion cannot come from any other process. It also differs from a shock propagation inside an NS (without combustion). For an axisymmetric star, the non-zero components of the gravitational wave field is \cite{zwerger,dimmelheimer}

\begin{equation}
h_{\theta \theta}^{TT} =\frac{1}{8} \sqrt{\frac{15}{\pi}} \sin^{2}\theta \frac{A_{20}^{E2}}{r}
\end{equation}
\begin{equation}
h_{\phi \phi}^{TT} =-h_{\theta \theta}^{TT}=h_{+}
\end{equation}
where $ \theta$ is the angle between the symmetry axis and the line of sight of the observer.

\begin{equation}
A_{20}^{E2}= \frac{d^{2}}{dt^{2}} \left( k \int \rho \left( \frac{3}{2} z^{2}-\frac{1}{2} \right)r^{4} dr dz \right)
\end{equation}
$$z=\cos{\theta}$$ and $$k=\frac{16 \pi^{3/2}}{\sqrt{15}}$$

\begin{figure} 
\includegraphics[width = 3.5in]{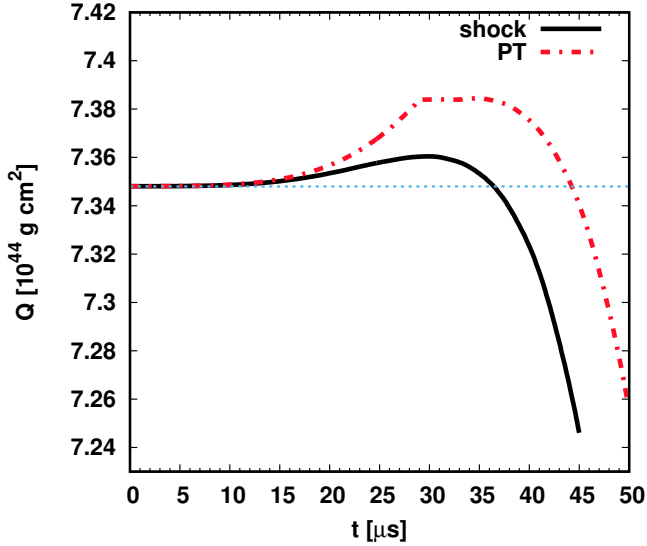}
\caption{The change in the Quadruple moment of the star as a function of time is shown in the figure. The change in Q is plotted when a shock wave is passing along the star and also for a star where there is combustion. The abrupt change in the quadrupole moment at 30 $\mu s$ is because at that instant combustion of Nm to QM stops and it
converts to a simple shock which propagates outwards.}
\label{quad}
\end{figure}

\begin{figure} 
\includegraphics[width = 3.5in]{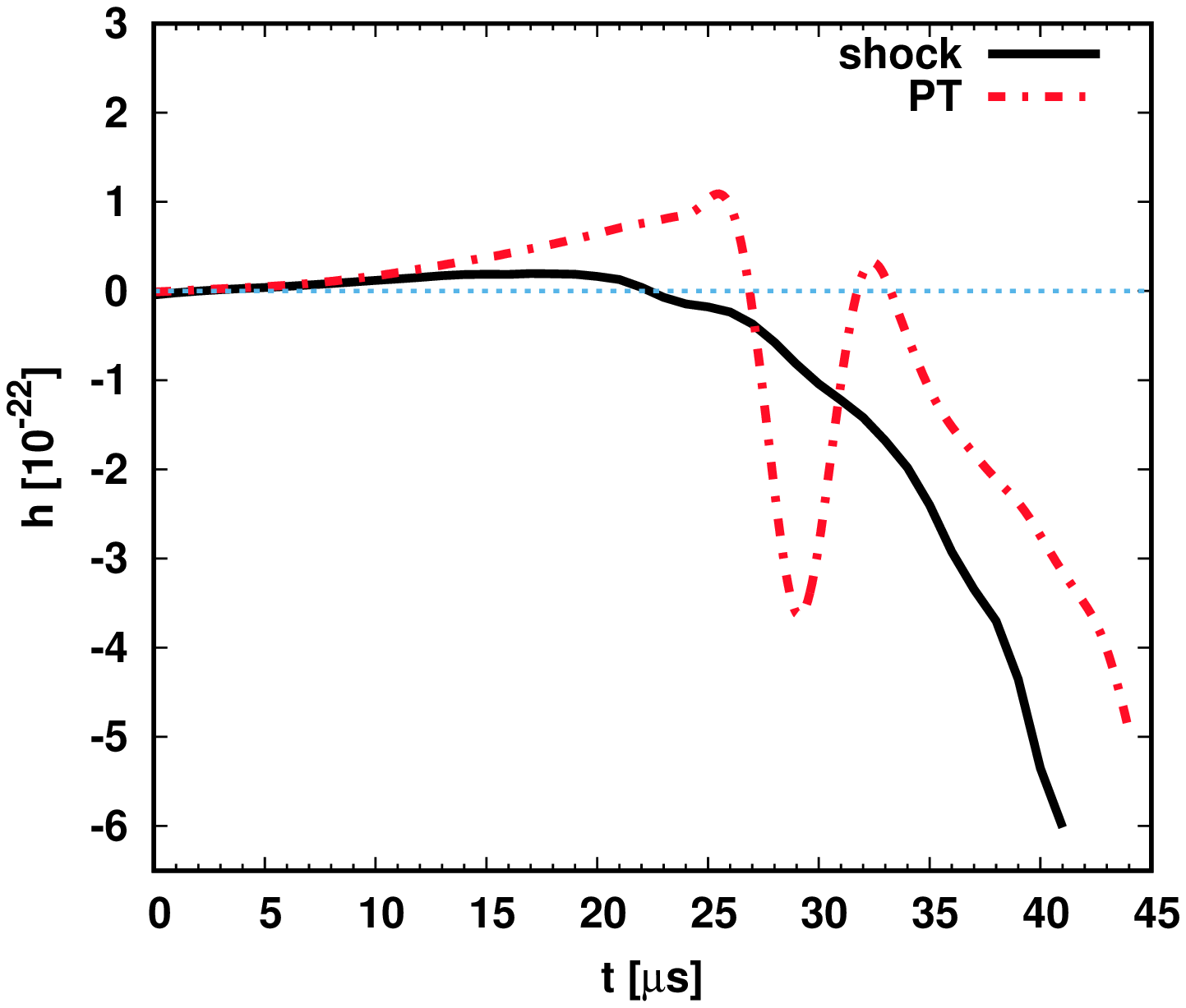}
\caption{The strain of the GW is shown as it evolves with time. Similar to the previous figure the strain is plotted for a star in which a simple shock wave passes and also for a 
star which suffers PT. The kink and rise of the strain of a combusting star occur when the combustion ends, and it converts to a shock wave. This occurs due to a sudden change in the 
quadrupole moment of the star.}
\label{strain}
\end{figure}

The above formulae are used for finding the GW emerging from supernova collapse and due to micro-collapse resulting from the aftermath of phase transition. In our recent work, we have modeled the dynamical PT process in NS, and it was observed that the configuration of a neutron star, like the density profile and pressure profiles, exhibited significant change during this process. In this work, we evaluate the quadrupole moment during the dynamical PT process and obtain the GW amplitude and frequency, which eventually sheds light on observation possibility in GW detectors. We present the detailed steps to perform the calculation of $h_{\theta \theta}^{TT}$ due to the phase transition. 

At $t=0$, we take the originating location of conversion font to be $r_{i}$ a distance very close to the center of the star. Using hydrodynamics simulation after a small time  interval (say $dt=10^{-7}$ sec) we evaluate the location ($r_{f}$) reached by conversion font. Once we know  $r_{f}$ for the first time step ($dt$), we obtain the density profile generated, wherein star is composed of quark matter EoS and hadronic matter EoS. We integrate the density profile to evaluate the corresponding gravitational wave amplitude $h_{\theta \theta}^{TT}$. Now this $r_{f}$ reached at the end of the first-time step will serve as $r_{i}$ for second-time step; we evaluate the final location reached by conversion font using hydrodynamics simulation at the end second-time step. We repeatedly find the $r_{f}$ and $h_{+}$ for every time step till the conversion font reach the surface of the star.  Contrary to PTNS where the 1-dimensional problem was solved, we extended our present calculation to 2-dimensions ($r$, $\theta$) using the RNS code.

In fig 8 we plot the quadrupole moment changes as a function of time. We have also shown a curve in which the quadrupole moment of a star changes when a shock propagates through a star. The quadrupole moment rises with time smoothly for a shock wave. However, the situation is quite different for combustion. Initially, it follows the shock curve but starts to deviate after some time. The combustion quadrupole moment curve lies well above the shock curve. Also after $30 \mu s$ (when it reaches a distance of 6 km) the curve first becomes flat and then dips as the shock reaches the outer layer of the star. The combustion stops at the point where the curve becomes flat and only the shock wave propagates out to the surface. Such a change in the quadrupole moment is likely to have some effect on the GW signal and can be very important for the detection of a particular template for such a scenario. 

Next, in fig 9 we plot the amplitude or the GW strain as a function of time, this involves evaluation double derivative of quadrupole moment and it is sensitive to the choice of time interval $dt$. We use the Savitzky-Golay \cite{savitzky}  filter to perform differentiation, smoothing the data by choosing appropriate window size such that it diminishes variations in data in $10^{-7}$ s, and differentiation is carried out for variations in quadrupole moment which are above $10^{-6}$s. This smoothing of variation suppresses the numerical noise in gravitational wave strain. We have shown curves for the shock propagation and combustion scenario. The shock wave strain first grows with time and then falls smoothly without having any oscillating or periodic behavior. However, the strain amplitude for the combustion of NM to QM is quite different. During the initial burning, the strain increases smoothly; however, at the point of the wave changing from combustion to shock, there is an oscillation of some sort in the signal. This signal is not present in the simple shock propagation and it a well-defined marker for the PT. This marker is particular to the combustion of NM to QM and finally in the formation of a hybrid star. The conversion of the wavefront from combustion to shock is very significant in giving rise to this exact signal. However, if the burning or the PT propagates to the surface of the star such signal would be not present. Such PT scenario is likely to have some different signal. Thereby, looking at the signal coming from an NS, we can likely to tell the difference between whether an NS has converted to a SS or an HS.

To obtain the power spectra, we adopt the general procedure to find power spectra of any time-domain signal $h(t)$ with $ 0 \leq t \leq T$. The signal has is a function of time and is descerete, where $n$ values of $h(t)$ is $h_{k}$ at intervals $t = k \times \Delta t$ where $n$ is an integer. The discrete fourier transform of signal is \cite{press,dimmelheimer},

\begin{equation}
\tilde{h}_{k}= \sum_{n=0}^{N-1} h_{n}  \exp{ \left( \frac{2 \pi i n k}{N} \right)}
\end{equation}

and the corresponding frequency is 
$$ \omega = \frac{2 \pi k}{T} .$$
The power spectrum is given by,
\begin{equation}
P(\omega)=\frac{|\tilde{h}_{k}|^{2}}{N^{2}}
\end{equation}
The amplitude spetrum is given by,
$$h(\omega)= \sqrt{P(\omega)}$$.

The plot of $P(\omega)$ vs. $\omega$ gives the power spectrum. It contains information about power content in different frequencies of the signal. For better prespective we plot the amplitude spectrum which is square root of power spectrum.

\begin{figure} 
\includegraphics[width = 3.5in]{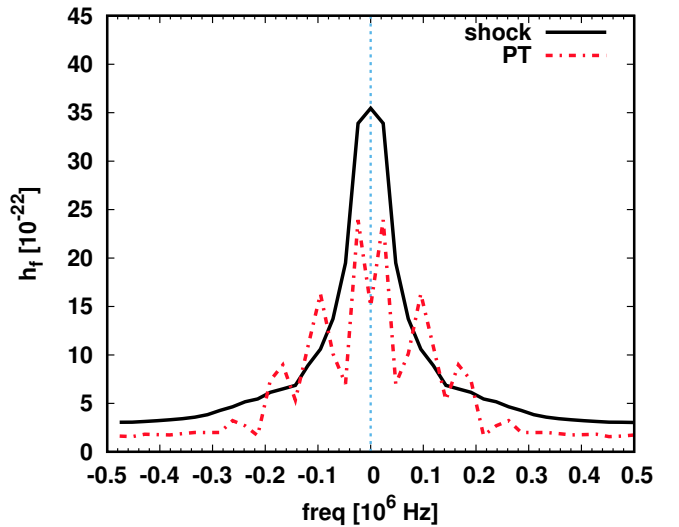}
\caption{ The amplitude spectrcal density is shown as a function of frequency.}
\label{power}
\end{figure}
 We plot the amplitude spectrum of the GW emission of conversion of NS to HS in fig 10. For portion of signal from $t=0\mu s$ to $t=41\mu s$ is considered and with a sampling time of $t=10^{-6}s$, the discrete Fourier transform is carried out. The sampling frequency $10^{6}$ Hz implies that the Nyquist frequency is $5 \times 10^{5}$ Hz, that is characteristic periodicities/ quasi-periodicities upto  $5 \times 10^{5}$ Hz can be observed if present in gravitational wave signal. There are three peaks appearing at $20$ KHz, $100$ KHz and $170$ KHz. The peaks appearing at such high frequency is also a unique signal of PT of NS to HS. We have only shown the power spectrum of the combustion scenario and not that of a shock. A pure shock does not show any peak. The peak of the power spectrum at such high frequency is a bit of challenge for GW telescopes. The usual telescopes work at a frequency of few HZ to a thousand Hz (KHz). New telescopes are increasing the frequency range to few KHz. However, such a GW signal has its first peak at tens of KHz, and the other peak occurs at hundred KHz and above. 
 
\begin{figure} 
\includegraphics[width =3.5in,height=2.8in]{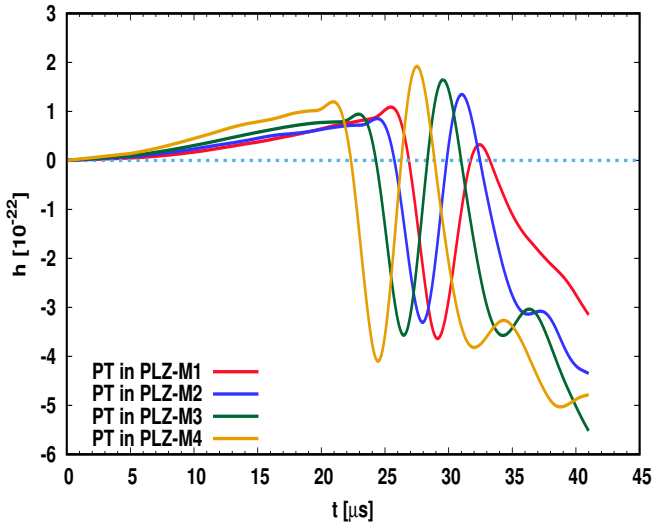}
\caption{The strain of the GW is shown as it evolves with time.The kink and rise of the strain of a combusting star occur when the combustion ends, and it converts to a shock wave. This occurs due to a sudden change in the quadrupole moment of the star.}
\label{strain2}
\end{figure}

We perform our simulation for different mass models, the different mass models considered and features of GW signature seen for PT in these stars listed in table 1. The gravitational wave strain for different neutron stars undergoing PT is shown in fig. \ref{strain2}, the PT front gets converted to shock front at 6km which results in a unique signature(a hump) which can be seen. The hump-like signature is seen to shift for different stars, the hump-like signature shifts towards left for massive NS , this is due to PT ending in different stars at different times. The corresponding amplitude spectral density for these signals is represented in figure 11, the peaks indicate the characteristic frequencies present in the GW signals.

\begin{figure} 
\includegraphics[width =3.5in,height=2.8in]{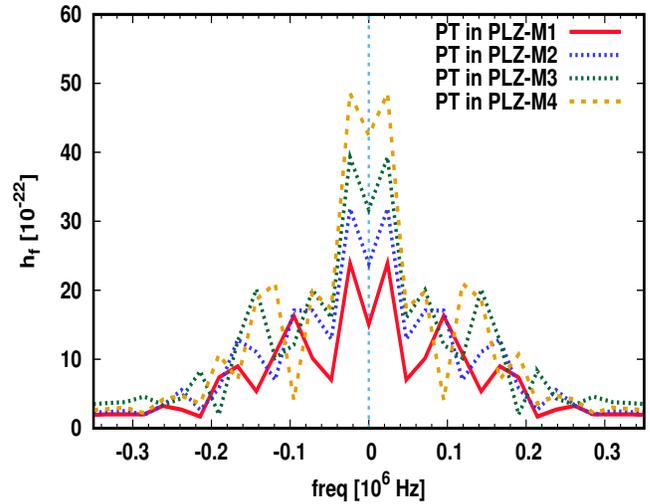}
\caption{Amplitude spectral density for GW produced from PT in various models.}
\label{power-all}
\end{figure}

Neutron star produces gravitational waves in different bands of frequency, some continuous and some short-lived. The continuous signals from neutron star are produced due to its rotation and deformed structure, these signals are usually monochromatic. When PT will happen in such a system two short-lived signals will emerge, one from dynamics of PT and other from micro-collapse. And eventually, the newly formed hybrid star too will emit continuous monochromatic GW. We give the overall picture of the gravitational wave signal coming from an NS, PT and its aftermath in fig 13 and table 2. Initially, the GW amplitude is very low ($10^{-26}$) and is generated by the rotation of the star. The star rotation is kept at $477$ Hz. The rotation of the star is periodic, and we get a sinusoidal curve. The zero of time is arbitrary and can be thought as the time we first start to observe the NS. Then we assume that at $t=2.60$ ms the PT starts, and the NM is combusting to QM. The strain is now of considerable strength ($10^{-22}$). At about 2.63 ms there is a sharp change in the strain (combustion ends, shock continues inside the star). This continues till few more $\mu s$ and then the collapse happens (along with the 2-f to 3-f conversion). The strain is considerable but is weaker than the NM to 2-f combustion signal. This would continue for a few more milliseconds till the central density of the star settles down to a stable HS. Finally, in the 4th panel, the strain (again of amplitude $10^{-26}$) of the HS is shown once the star has settled down. This GW is again coming due to the rotation of the HS. There is a slight change in the amplitude and frequency of the initial NS and final HS as the HS is more compact. The ellipticity of isolated radio pulsars is seen to be between $10^{-4} - 10^{-6}$. The ellipticity of quark/strange stars is thought to be higher than neutron stars, hence we have chosen $10^{-4}$ as ellipticity of neutron star and $10^-{3}$ as ellipticity of hybrid star \cite{aasiel,lansky}.

\begin{figure*} 
\includegraphics[width = 7in]{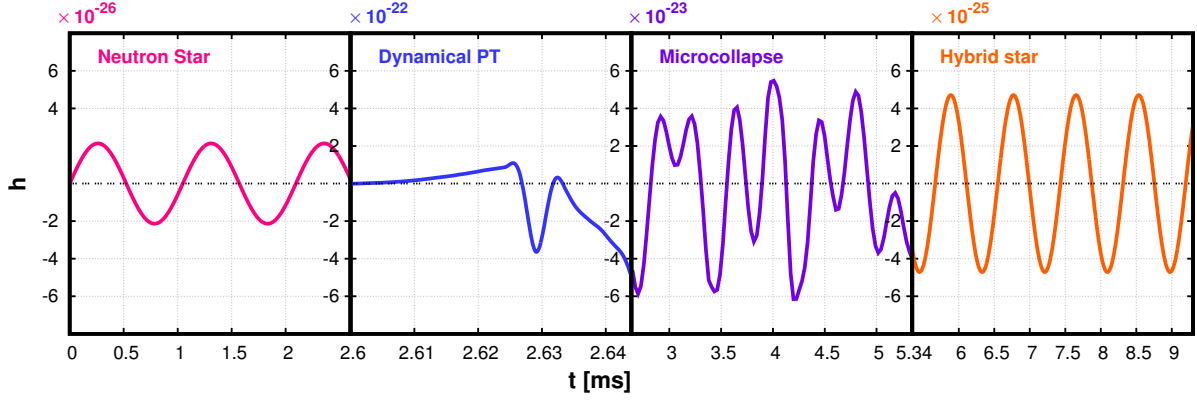}
\caption{We show the strain for the fours stage of combusting NS. The 1st panel shows the GW strain due to the rotation of NS. The second panel shows the GW strain of NM 
to QM combustion. The third panel shows the collapse of a star. Finally, the fourth panel shows the GW from a rotating HS. Note that the strain of different panels are of 
different strength.}
\label{tot-wave}
\end{figure*}

%\begin{widetext}

\begin{table*}
\centering
\caption{Properties of initial NS and final 3-flav hybrid star}
\begin{tabular}{ccccccccc}
\hline 
\hline
Model &  $\rho_{c}$ &  $\omega$  & $M$ & $r_{e}/r_{p}$ &  $T/W$ & I & $\epsilon$   \\ 
     &   ($10^{14}$  g/cc) & ($10^{4}$ Hz) & ( $M_{sun}$) &  &  ($10^{-2}$)  & $10^{45}$ g cm$^{2}$ &(choosen) \\
\hline
NS (PLZ-M1) &5.0& 0.300& 1.50& 1.28871& 2.73182& 2.22& $10^{-4}$  \\
HS & 5.37 &  0.356 &   1.43 &  1.16406 &  3.33161 &  1.87882  & $10^{-3}$  \\
\hline
\end{tabular}
\end{table*}
%\end{widetext}

\section{Summary and conclusion}
In this work, we have done a simulation of an axisymmetric star undergoing combustion from NS to QS. We have mimicked a 2-d simulation by employing the GR1D code in  65 direction of the star (the star is sliced into 65 direction from and the angle between 0 to $\pi/2$. This gives the single quadrant of the star. All the other three quadrants are a replica of this quadrant. Therefore, evolving a single quadrant, we get the information of the whole star. The conversion starts at some point near the center of the star. As the conversion propagates outward, it converts NM to QM. As the star is now axisymmetric and has an oblate spheroid shape, the density distribution along the polar and equatorial direction are not the same. The combustion therefore, proceeds with different velocity along the different direction of the star. We evolve this 65 profiles(each corresponding to a $\theta$) one by one. The output of hydrodynamics simulation gives the value of  
density $\rho (r, \theta_{fixed})$, pressure $p (r, \theta_{fixed})$ and velocity $v (r, \theta_{fixed})$ at each point in the system at any time $t$ along a given  direction. By combining all these 65 profiles along 65 directions, we get the complete information of star.
The combustion starts at a distance of $0.5$ km and continues till $6$ km. Till this point the combustion wave continues, and NM suffers a PT. Beyond this distance of the star, the NM is much stable than QM, and there is no PT. Therefore beyond that point, we have a shock wave propagating to the surface of the star. To calculate the prospect of GW signal due to the combustion of NM to 2-f QM, we have done a relativistic hydrodynamic simulation of the combustion process neglecting gravity. The gravity is not likely to influence the process significantly due to fast burning. The gravity is expected to have a significant effect on the 2nd step process (2-f QM to 3-F QM) where the star suffers micro-collapse.

As the combustion wave propagates outwards, the internal dynamics of the star changes very rapidly. This changes the overall quadrupole moment of the star which happens very fast.As the quadrupole moment of the star changes, the star emits strong GW emission with the strain of the order of $10^{-22}- 10^{-23}$ depending on the distance of the NS. The Quadrupole moment of the NS changes when the combustion wave ends, and the shock wave begins. In other words, as the PT ceases to take place, there is a change in the quadrupole moment of the star. The gravitational wave amplitude or the strain of the GW depends strongly on the change in quadrupole moment. Therefore, the point where the combustion of NM to QM stops there is a sharp change in the amplitude of the strain. Such a change in quadrupole moment and strain is not seen for ordinary shock wave propagation. This result is very typical of combustion of NM to QM. This marker of a sudden change in the GW strain is very typical of the formation of HS. This type of signal is not present even for the conversion of NS to SS, where the combustion front goes to the surface of the star. This marker can be used to point the difference between an NS to HS conversion and NS to SS conversion. The only problem with the detection of such type of GW signals coming from PT scenario is that the peak in the power spectrum lies in upper KHz to MHz range. The appearance of the peaks at the high-frequency range is also a typical signal of such NS to HS conversion. A simple shock wave propagation does not have such two peaks in the power spectrum.  Finally, we have shown the overall picture of such PT from NS to HS. A rotating star emits very low amplitude GW signal beyond our detection capability. As the combustion from NM to QM starts in the star the strain amplitude peaks up and increases continuously. The time when the burning stops a small oscillation in the wave amplitude generates. The combustion lasts only for tens of microseconds. Then the star settles down to a more stable 3-f QM, and the star suffers a collapse. This results in an oscillation of the central density of the star. This process continues for a few milliseconds and the GW strain at this time is of the order of $10^{-23}$. After the HS settles down, it again starts to emit very low amplitude GW signals due to rotation which is hard to detect.

In this work, we perform a real dynamical evolution of PT in NS which results in the formation of an HS. We have shown that such PT has a very unique signal which is very different from any other astrophysical signals. Such signals can help us in confirming whether PT in NS is possible or not. It would also help us in determining whether a PT has resulted in the formation of SS or an HS. A hybrid star has features which are very similar to that of NS, and it is very hards to distinguish them. However, we have shown that the PT scenario can distinguish between the formation of SS and HS. We should mention here that we have done the dynamical calculation of the 1st process of PT, that is NM to 2-f QM dynamics. The 2-f to 3-f QM formation dynamics is still to be done. The 2nd process dynamics time scale is similar to the timescale of the star collapse. In our future work, we will be trying to address the dynamics of the second process. We also mention that we have not taken gravity into account as we believe that the effect of gravity in such a fast process is not very huge.
However, to be through presently, we are trying to incorporate gravity in our calculation.

\acknowledgments
The author RM is grateful to the SERB, Govt. of India for monetary support in the form of Ramanujan Fellowship (SB/S2/RJN-061/2015) and Early Career Research Award (ECR/2016/000161). RP would like to acknowledge the fianancial support in form of INSPIRE fellowship provided by DST, India. RM and RP would also like to thank IISER Bhopal for providing all the research and infrastructure facilities.

\end{document}